\documentclass[twocolumn,pra,notitlepage]{revtex4-1}
\usepackage{amsmath}
\usepackage{graphicx}
\usepackage{wrapfig}
\usepackage{enumerate}
\usepackage{float}
\usepackage{standalone}
\usepackage{lineno}
\usepackage{subcaption}

\begin{document}

\title{Narrow-linewidth Fano microcavities with resonant subwavelength grating mirror}

\author{Trishala Mitra$^1$, Gurpreet Singh$^1$, Ali Akbar Darki$^1$, S{\o}ren Peder Madsen$^2$and Aur\'{e}lien Dantan$^1$}\email{dantan@phys.au.dk} 

\address{$^1$Department of Physics and Astronomy, Aarhus University, DK-8000 Aarhus C, Denmark\\
$^2$Department of Mechanical and Production Engineering, Aarhus University, DK-8000, Aarhus C, Denmark}




\begin{abstract}
We report on the theoretical and experimental investigations of optical microcavities consisting in the plane-plane arrangement of a broadband high-reflectivity mirror and a suspended one-dimensional grating mirror possessing a high-quality factor Fano resonance. By varying the length of these cavities from the millimeter to the few-micron range, we observe at short lengths the reduction of the spectral linewidth predicted to occur for such a Fano cavity as compared to a conventional broadband mirror cavity with the same length and internal losses. Such narrow linewidth and small modevolume microcavities with high-mechanical quality ultrathin mirrors will be attractive for a wide range of applications within optomechanics and sensing.
\end{abstract}

\date{\today}

\maketitle


\section{Introduction}

Small modevolume microcavities are a key ingredient in numerous applications in which enhanced light-matter interactions are desirable, e.g. optical sensing, nonlinear optics and lasers, cavity quantum electrodynamics or cavity optomechanics~\cite{Vahala2003}. For most of these applications the spectral selectivity of the cavity is an important figure of merit. For sensing applications directly exploiting the mechanics---as in gas pressure sensing~\cite{Naserbakht2019,AlSumaidae2021,Hornig2022,Salimi2023}---or for cavity optomechanics investigations of the interaction between optical and mechanical modes~\cite{Aspelmeyer2014}, the use of Fabry-Perot--type cavities using ultrathin mirrors with excellent mechanical properties~\cite{Kemiktarak2012,Bui2012,Kemiktarak2012a,Norte2016,Reinhardt2016,Chen2017,Zhou2023,Enzian2023} may be advantageous in terms of optical access and of independently optimizing the optical and mechanical response of the system. However, for a fixed loss level, the reduction in the cavity length of such a linear Fabry-Perot cavity traditionally implies an increase in its linewidth and decrease in spectral selectivity.

\begin{figure}[h]
\includegraphics[width=\columnwidth]{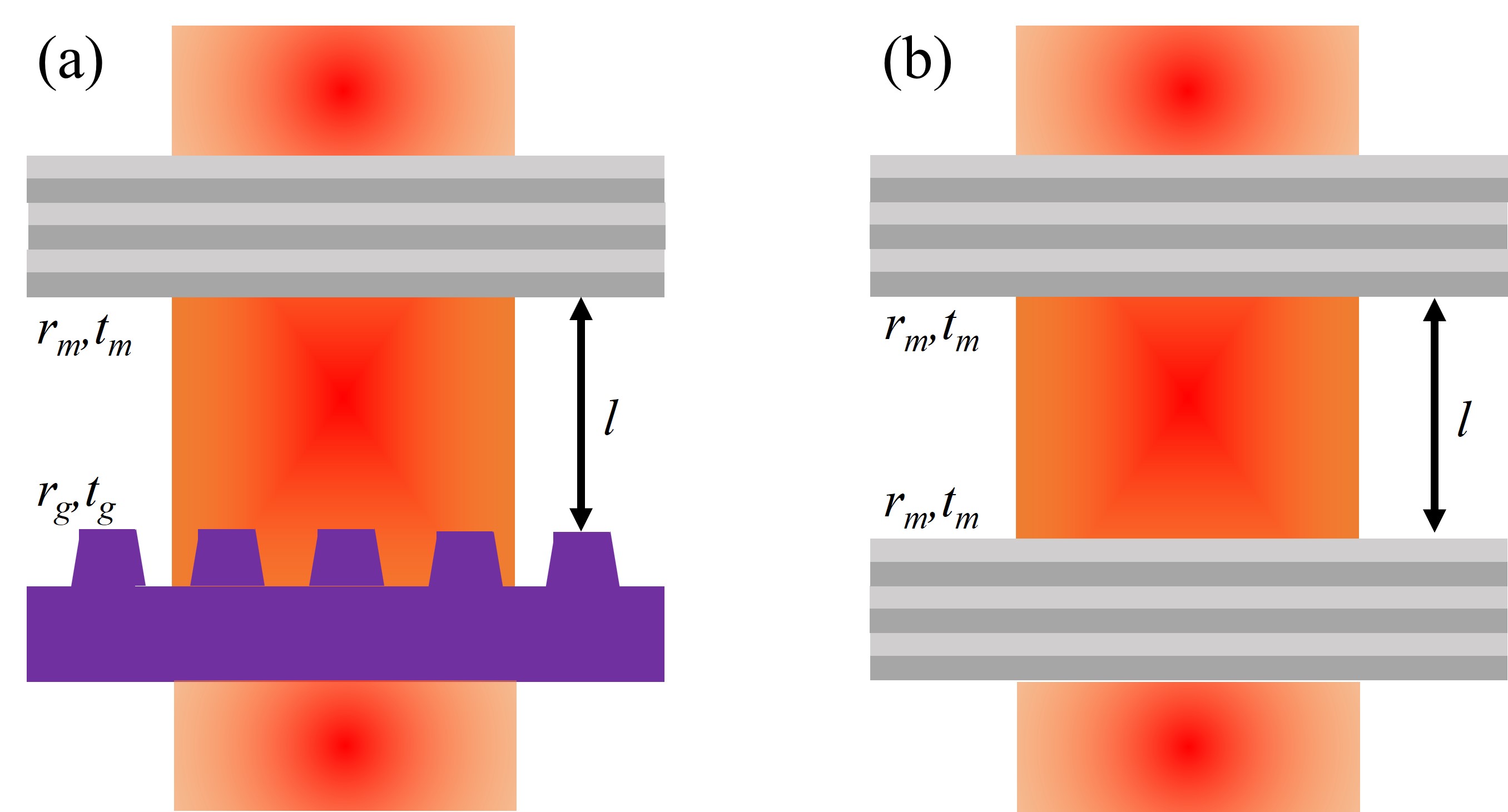}
\caption{(a) Schematic Fano cavity consisting of a highly reflective Bragg mirror with wavelength-independent transmission/reflection coefficients $r_m$/$t_m$ and of a resonant subwavelength grating mirror with strongly wavelength-dependent transmission/reflection coefficients $r_g$/$t_g$. (b) Corresponding broadband mirror cavity with equal length.}
\label{fig:Fano_broadband}
\end{figure}

Recently, we proposed to realize narrow-linewidth ultrashort microcavities using a structured, ultrathin mirror suspended on top of a conventional broadband high-reflectivity mirror~\cite{Naesby2018}. This "Fano cavity" exploits the fact that a suitably structured ultrathin mirror may exhibit a narrow optical Fano resonance arising from the coupling of the out-of-plane light with a guided-mode in the structure~\cite{Wang1993,Fan2003,Limonov2017,Quaranta2018}. When the cavity resonance matches that of the Fano mirror and for short cavity length and sufficiently high-Q resonances, it was predicted that such cavities may exhibit a linewidth which can be substantially narrower than both the Fano mirror resonance and the linewidth of a broadband cavity with same length and internal losses~\cite{Naesby2018} (see Fig.~\ref{fig:Fano_broadband}). Such narrow-linewidth Fano cavities have been predicted to give rise to interesting applications for cavity optomechanics~\cite{Cernotik2019,Fitzgerald2021,Manjeshwar2023,Peralle2023}, and nonconventional optomechanics interactions have been very recently observed with semiconductor photonic crystal membranes suspended on top of a Bragg reflector~\cite{Manjeshwar2023}. Note also that this approach, based on the resonant coupling with a guided-mode, is different than minimizing metasurface scattering losses by careful phase design as in Ref.~\cite{Ossiander2023}.

We report here on the realization of Fano microcavities using a resonant subwavelength grating mirror made on highly-pretensioned silicon nitride thin films and suspended on top of a conventional broadband high-reflectivity mirror. We investigate the transmission properties of these cavities on the basis of both an analytical "plane-wave" model as well as finite element simulations taking into account the finite transverse size of the grating/cavity and the Gaussian nature of the incident beam. Experimentally, by varying the length of these cavities from the millimeter to the few-micron range, we are able to observe the linewidth narrowing predicted in Ref.~\cite{Naesby2018} and to demonstrate a significant linewidth reduction as compared to the corresponding broadband mirror cavities.

The intrinsic low loss of such subwavlength grating membranes combined with their excellent mechanical properties~\cite{Darki2022} make the investigated Fano cavities promising for a wide range of optomechanical applications, e.g. the realization of low-light level optical bistable and nonreciprocal optomechanical cavities~\cite{Sang2022,Xu2022,Zhou2023,Enzian2023} or the investigations of multiple-membrane cavity optomechanics~\cite{Xuereb2012,Xuereb2014,Nair2017,Gartner2018,Piergentili2018,Wei2019,Manjeshwar2020,Yang2020b} in the strong coupling regime, but also potentially for optical sensing~\cite{Naserbakht2019,AlSumaidae2021,Hornig2022,Salimi2023} and Fano lasers~\cite{Mork2014,Yu2017}.

The paper is organized as follows: in Sec.~\ref{sec:theory} we present a model for "plane-wave" Fano cavities with a lossy resonant subwavelength grating, which allows for the generic comparison of Fano and broadband mirror cavities as well as sets the basis for the analysis of the experiments presented in the subsequent section. Section~\ref{sec:theory} also presents the results of finite element simulations of both types of cavities taking into account their finite transverse size and the Gaussian nature of the incident beam. In Sec.~\ref{sec:experiment} the experimental investigations of Fano cavities using moderate-Q or high-Q resonant subwavelength gratings are reported. Section~\ref{sec:conclusion} concludes the paper.


\section{Fano cavity with subwavelength grating mirror}
\label{sec:theory}

\subsection{Subwavelength grating model}

\subsubsection{Lossless grating}

We first consider an infinite subwavelength grating, surrounded by air on both sides and illuminated at normal incidence by a linearly polarized monochromatic plane wave, with a wavelength close to a guided-mode resonance wavelength. We describe the normal incidence transmission and reflection amplitude coefficients of a lossless resonant grating by generalizing the model of Fan and Joannopoulos~\cite{Fan2003}
\begin{equation}
t_g=t_d+\frac{a}{k-k_1+i\gamma}\hspace{0.5cm}\textrm{and}\hspace{0.5cm} r_g=r_d+\frac{b}{k-k_1+i\gamma},
\end{equation}
where $t_d$ and $r_d$ are the direct transmission and reflection coefficients of the slab far from resonance, $k=2\pi/\lambda$ is the wavenumber, $k_1=2\pi/\lambda_1$ is the guided-mode resonant wavenumber and $\gamma$ determines the width of the guided-mode resonance. $a$ and $b$ are complex coefficients describing the interference between the direct transmitted/reflected waves and the guided mode. 

In Ref.~\cite{Fan2003} $a$ and $b$ are assumed to be equal, and the energy conservation relations
\begin{equation}
|t_g|^2+|r_g|^2=1\hspace{0.5cm}\textrm{and}\hspace{0.5cm}|t_d|^2+|r_d|^2=1
\label{eq:conservation}
\end{equation}
lead to $a=b=-i\gamma(t_d+r_d)$. This yields in turn a grating transmission amplitude which can be put under the form
\begin{align}
t_g=t_d\frac{k-k_0}{k-k_1+i\gamma},
\end{align}
where $k_0=k_1-i\gamma(r_d/t_d)$ is the zero-transmission/unity-reflection wavenumber. One can also show that unity-transmission is achieved at $k_t=k_1+i\gamma(t_d/r_d)$.\\

However, in general---when the grating structure is not vertically symmetric~\cite{Popov1986}---$a$ and $b$ need not be equal~\cite{Bykov2015,Darki2021,Parthenopoulos2021}. Assuming for simplicity $t_d$ and $r_d$ real, the energy conservation relations (\ref{eq:conservation}) impose
\begin{align}
\label{eq:xb} & t_dx_a+r_dx_b=0,\\
\label{eq:yb} & x_a^2+y_a^2+x_b^2+y_b^2+2t_d\gamma y_a+2r_d\gamma y_b=0,
\end{align}
where $x_{a,b}$ and $y_{a,b}$ are $a$ and $b$'s real and imaginary parts, respectively. The grating transmission coefficient can now be put under the form
\begin{equation}
t_g=t_d\frac{k-k_0+i\beta}{k-k_1+i\gamma},
\label{eq:tg_general}
\end{equation}
where $k_0$ and $\beta$ are defined from $a=t_d(k_1-k_0+i\beta-i\gamma)$. The reflectivity coefficient can then be calculated from Eqs.~(\ref{eq:xb},\ref{eq:yb}). Non-zero transmission around $k_0$ and non-unity transmission around $k_t$ are then possible.

\subsubsection{Lossy grating}

To model guided-mode losses, one can modify the energy balance relation by adding a resonant loss term as
\begin{align}
|t_g|^2+|r_g|^2+\frac{c^2}{(k-k_1)^2+\gamma^2}=1,
\end{align}
where $c$ is related to the resonant loss level $L$ by $c^2=L((k_0-k_1)^2+\gamma^2)$. If one fits the transmission profile with Eq.~(\ref{eq:tg_general}) and if $L$ is known, the reflectivity coefficient can be found by solving
\begin{align}
\label{eq:xb_general} & t_dx_a+r_dx_b=0,\\
\label{eq:yb_general} & x_a^2+y_a^2+x_b^2+y_b^2+c^2+2t_d\gamma y_a+2r_d\gamma y_b=0.
\end{align}

\begin{figure}
\includegraphics[width=\columnwidth]{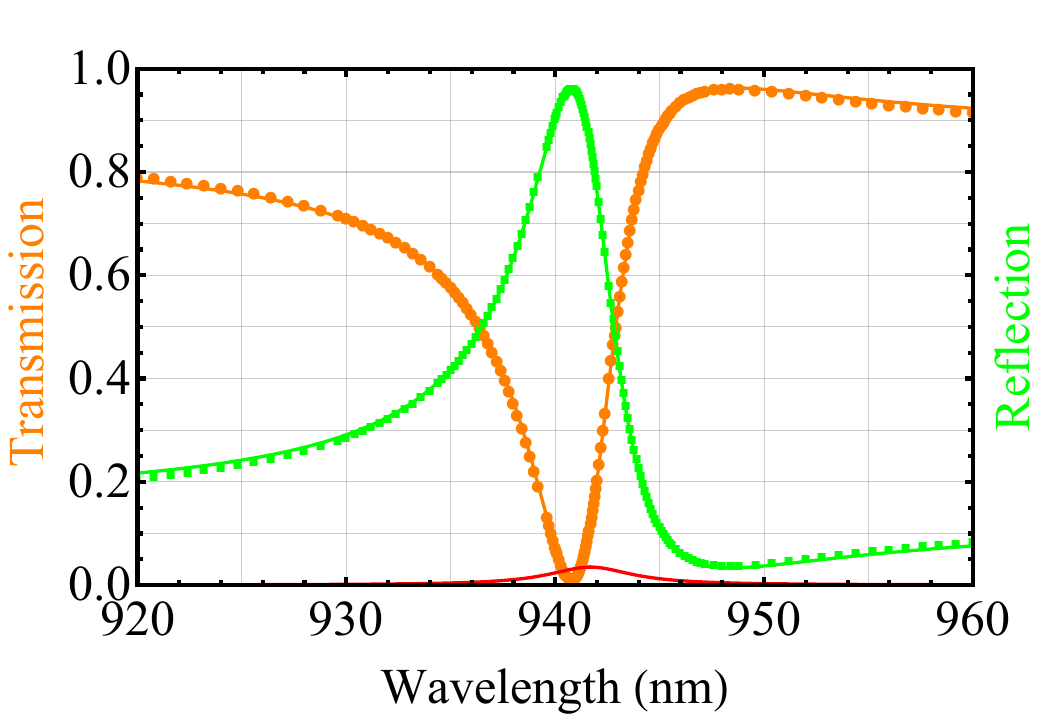}
\caption{Simulated transmission (orange circles) and reflection (green squares) spectra of a moderate Q, lossy grating. The orange and green lines show the resulting fit to the model (Eqs.~(\ref{eq:tg_general}), (\ref{eq:xb_general}) and (\ref{eq:yb_general})), and the red line shows the losses $1-|t_g|^2-|r_g|^2$, respectively. }
\label{fig:comsol_grating_only}
\end{figure}

To illustrate the accuracy of the model, Figure~\ref{fig:comsol_grating_only} shows simulated transmission and reflectivity spectra of a moderate Q grating with parameters similar to one of the gratings (grating $A$) investigated experimentally in Sec.~\ref{sec:experiment}. The grating has rectangular fingers with period 840 nm, width 400 nm, depth 150 nm for a total film thickness 200 nm and a refractive index of 2.0. In the Finite Element Method simulations using the COMSOL Multiphysics environment, the two-dimensional grating---infinite in the grating finger direction and with a transverse extension of 90 $\mu$m---is illuminated by a TM-polarized Gaussian beam with a waist (radius) of 30 $\mu$m. Collimation and finite-size effects~\cite{Magnusson1993,Jacob2000,Bendickson2001,ToftVandborg2021} result in a non-zero resonant transmission (0.9\%) and non-unity peak reflectivity (96\%) which can be evaluated by fitting the simulated data with the previous expressions. The best fit parameter results for the simulated data shown in Fig.~\ref{fig:comsol_grating_only} are $\lambda_0=940.8$ nm, $\lambda_1=941.7$ nm, $\gamma_\lambda=2.4$ nm, $|t_d|^2=85.6\%$, $\beta=1.9$ $\mu$m$^{-1}$ and $L=3.1\%$.

\subsection{Fano cavity model}

In order to model the Fano cavity transmission spectrum we make use of the previously determined wavelength-dependent grating transmission and reflection coefficients, $t_g$ and $r_g$, in the planar Fabry-Perot transmission function
\begin{align}
\mathcal{T}_\textrm{cav}=\left|\frac{t_mt_ge^{i\phi}}{1-r_mr_g e^{2i\phi}}\right|^2
\label{eq:Tcav}
\end{align}
where $t_m$ and $r_m$ denote the incoupling broadband mirror's transmission and reflection coefficients and $\phi=kl$, where $l$ is the cavity length. For a cavity resonance close to the grating zero-transmission resonance the cavity transmission is well-approximated by
\begin{align}
\mathcal{T}_\textrm{cav}\simeq\frac{A}{1+\left(\frac{\Delta}{1-\nu\Delta}\right)^2}+B,
\label{eq:cavitylinewidth}
\end{align}
where $A$ and $B$ are constants, $\Delta=(\lambda-\lambda_c)/\delta\lambda$ is the wavelength detuning (normalized by the HWHM $\delta\lambda$) from the cavity resonance $\lambda_c$ , and where $\nu$ is a constant determining the degree of asymmetry of the Fano transmission profile. One can show that, when $\lambda_c\simeq\lambda_0$, the cavity HWHM $\delta\lambda$ is approximately given by
\begin{align}
\delta\lambda\simeq\frac{1}{\frac{1}{\delta\lambda_c}+\frac{1}{\delta\lambda_g}},
\label{eq:deltalambda}
\end{align}
where
\begin{align}
\delta\lambda_c=\frac{\lambda_0^2}{8\pi l}(T_g+T_m+L)
\label{eq:deltalambda_b}
\end{align} is the HWHM of the broadband cavity and
\begin{align}
\delta\lambda_g=\frac{\gamma_\lambda}{2(1-r_d)}(T_g+T_m+L)
\label{eq:deltalambda_f}
\end{align}
is the HWHM of the Fano cavity in the Fano regime. $\gamma_\lambda=(\lambda_1^2/2\pi)\gamma$ represents the grating resonance HWHM in wavelength and $T_g=|t_g(\lambda_0)|^2$ and $T_m=|t_m|^2$ represent the intensity transmission levels of the grating and the mirror at resonance.

\begin{figure}
\includegraphics[width=\columnwidth]{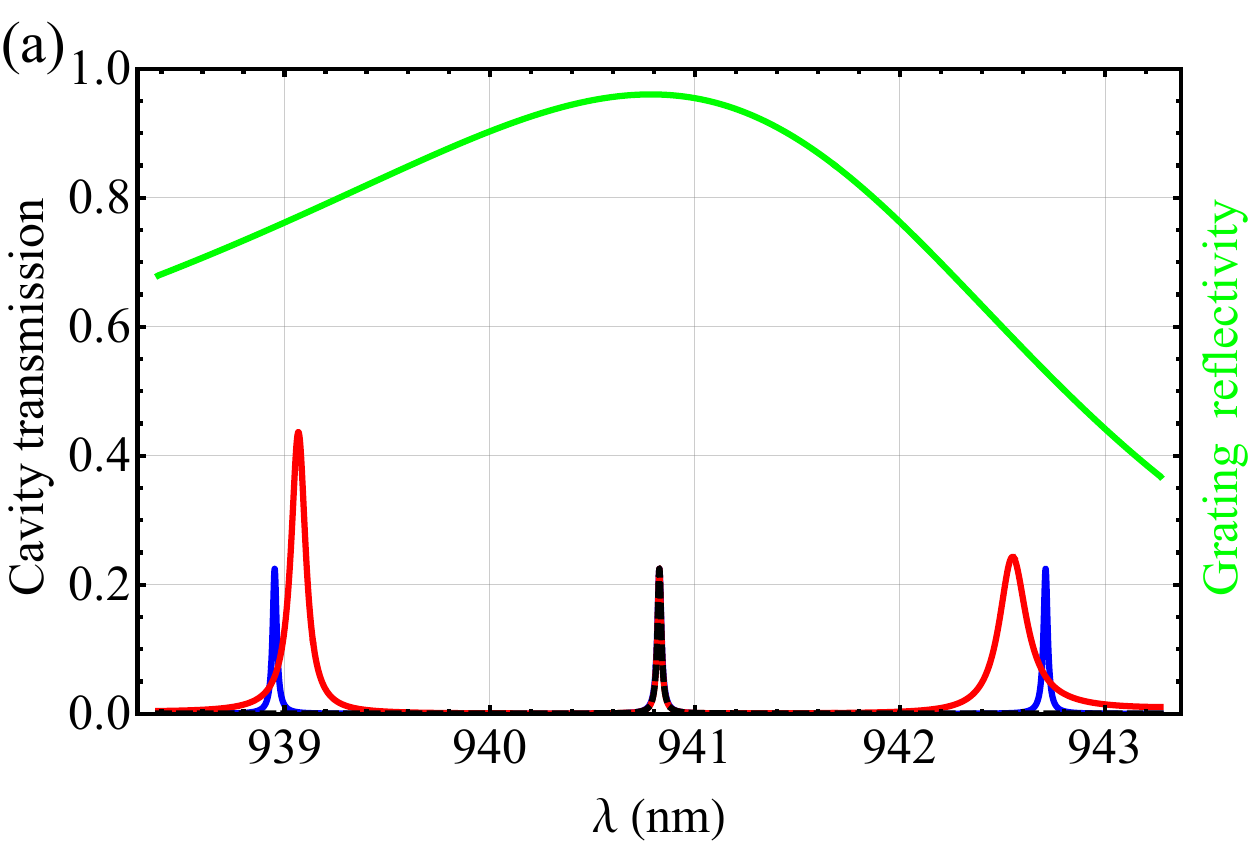}
\includegraphics[width=\columnwidth]{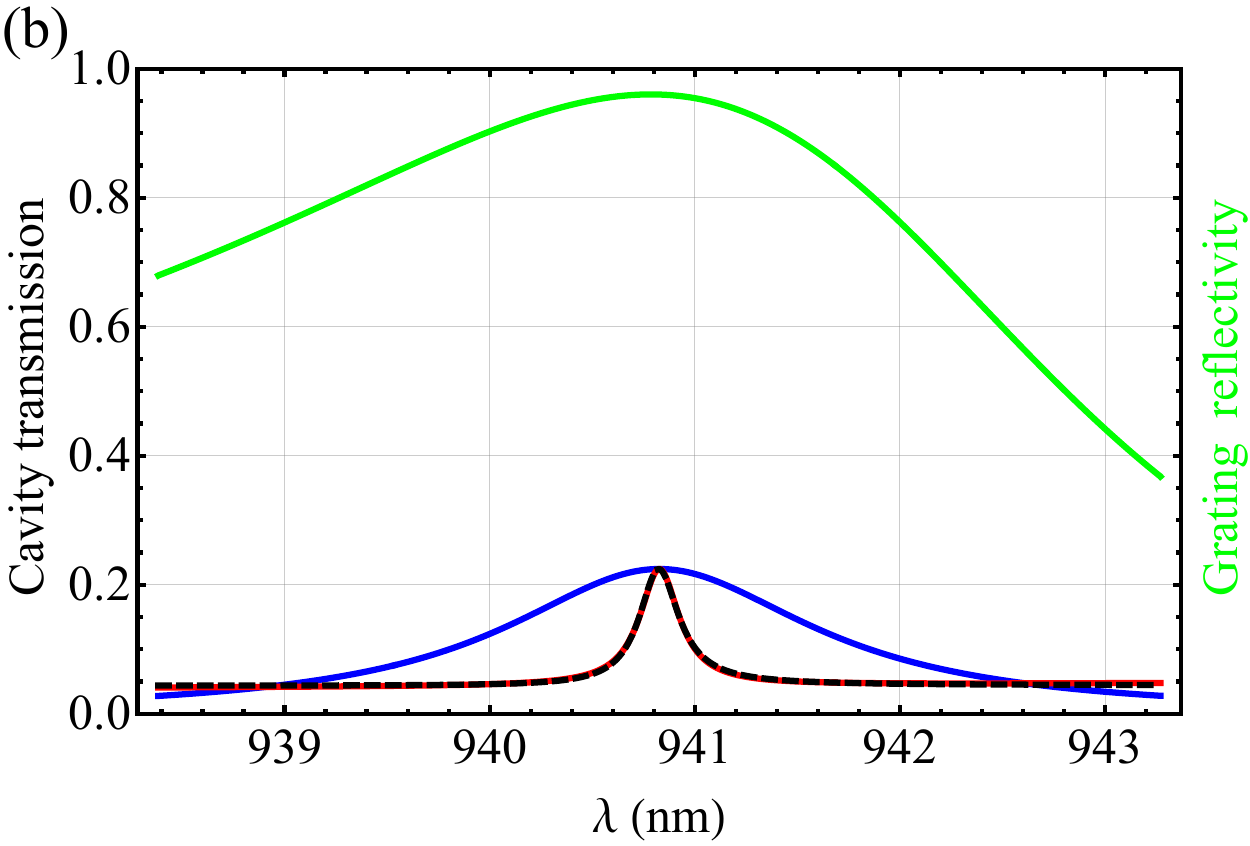}
\caption{Cavity transmission for a $l=235.5$ $\mu$m-long cavity (a) and $l=3.1$ $\mu$m-long cavity (b). The red curves show the transmission of a Fano cavity consisting of the grating simulated in the previous section and of a broadband 96\% reflective mirror. The black dashed curves show the result of a fit with the Fano cavity model. The blue curves show the transmission of the corresponding broadband cavity consisting of two 96\% reflective mirrors. The green curves show the reflectivity of the grating.}
\label{fig:cav}
\end{figure}

Figure~\ref{fig:cav} shows examples of cavity transmission spectra for the grating from the previous section ($T_g=0.9$\%, $L=3.1$\%, $\gamma_\lambda=2.4$ nm) and a broadband mirror with $T_m=4$\%. Two cavity lengths were chosen to illustrate the "long" ($l=235.5$ $\mu$m), standard cavity regime and the short ($l=3.1$ $\mu$m), Fano cavity regime. Fitting the resonance for the 235.5 $\mu$m-long cavity with Eq.~(\ref{eq:cavitylinewidth}) yields $\delta\lambda=11.8$ pm, as predicted from Eq.~(\ref{eq:deltalambda}) ($\delta\lambda_c=11.1$ pm, $\delta\lambda_g=157$ pm). For the 3.1 $\mu$m-long cavity, a fit with Eq.~(\ref{eq:cavitylinewidth}) gives $\delta\lambda=117$ pm, while Eq.~(\ref{eq:deltalambda}) predicts $\delta\lambda=134$ pm  ($\delta\lambda_c=900$ pm, $\delta\lambda_g=157$ pm). While Fig.~\ref{fig:cav}(a) illustrates the fact that the Fano cavity behaves as a broadband cavity in the long cavity regime, Fig.~\ref{fig:cav}(b) shows that, in the short cavity regime, the Fano cavity resonance can be much narrower than that of of the Fano mirror itself or of the corresponding broadband cavity.

\subsection{Comparison of Fano and broadband mirror cavities}
\label{sec:comparison}

To check the accuracy of the predictions of the previous model, as well as to assess the effects due to the Gaussian nature of the incident beam and the finite size of the grating, finite element method simulations were performed using the COMSOL Multiphysics environment in a two-dimensional model (the grating is assumed infinite in the grating finger direction). 

\begin{figure}[h]
\includegraphics[width=\columnwidth]{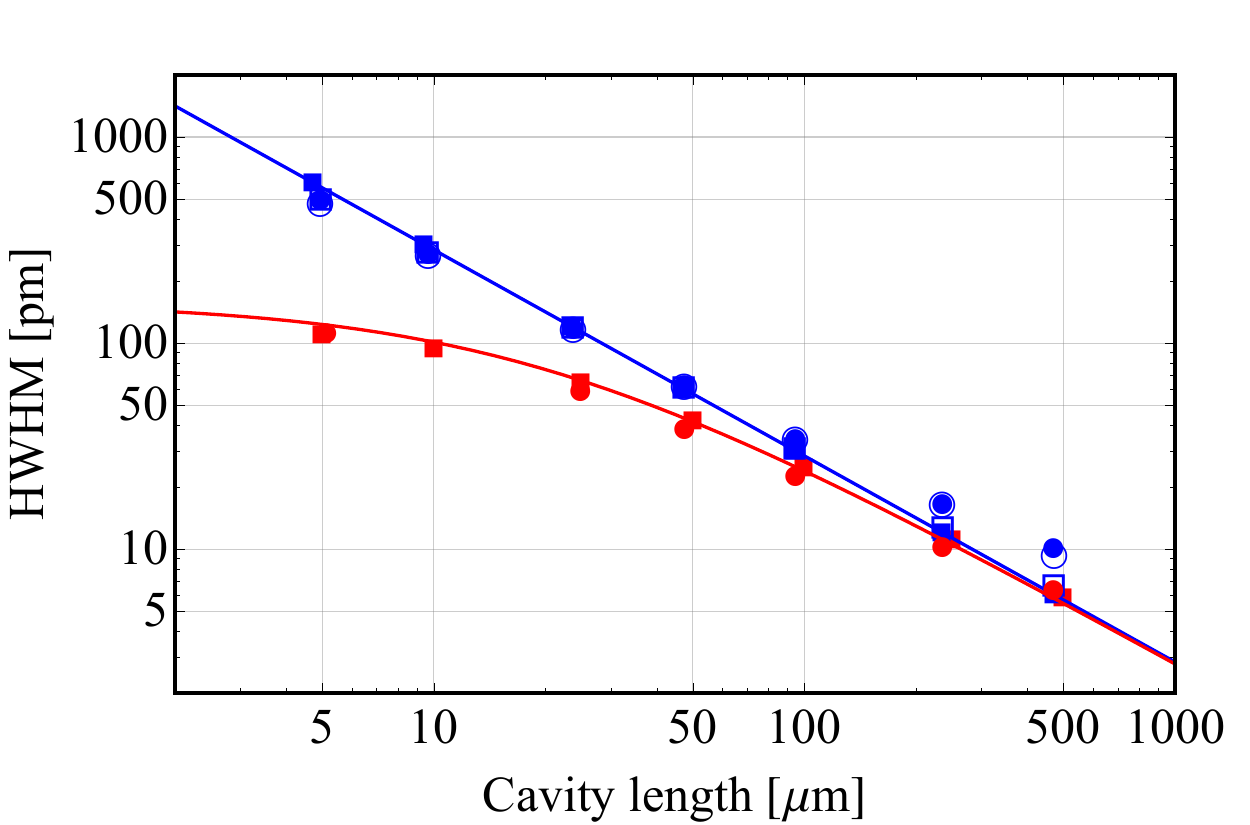}
\caption{Variations of cavity linewidth (HWHM) with cavity length of broadband mirror (blue) and Fano cavities (red) having 8\% internal losses and resonant at the grating resonance wavelength. The full squares show the results of infinite structure/plane wave simulations, while the full circles show the results of finite structures illuminated by a Gaussian beam with waist (see text for details). The blue and red lines show the analytical predictions of Eqs.~(\ref{eq:deltalambda_b}) and (\ref{eq:deltalambda}). In addition, the empty blue circles and squares indicate the predictions based on a spatiospectral decomposition of the incident Gaussian beam of the HWHM of a broadband mirror cavity illuminated by Gaussian beams with waist $w_0=30$ and 90 $\mu$m, respectively.}
\label{fig:comsol}
\end{figure}

In a first series of simulations the grating is assumed illuminated by an infinite plane wave and Floquet periodic boundary conditions are applied on each side of a single period (unit cell) of the grating. The high-reflectivity mirror is modeled as a Bragg grating structure consisting of three pairs of lossless dielectric media with refractive index 3.22 and 1.5. The thicknesses of the layers are chosen as to ensure a broadband reflectivity of 96\% around the target resonant wavelength $\lambda_0=941$ nm. The broadband mirror cavity is realized by considering two parallel such Bragg grating mirrors, while the Fano cavity consists of one Bragg mirror and the grating. In order to match the total cavity losses of 8\% of both cavities, the grating is made slightly absorptive by taking its refractive index to be $n=2+i(4.8\times 10^{-4})$. Figure~\ref{fig:comsol} shows the variation of the linewidth of the cavity for different cavity lengths chosen such that the cavity is resonant at $\lambda_0$, and for both the broadband mirror cavity (blue squares) and the Fano mirror cavity (red squares). The blue and red lines show the analytical predictions of Eqs.~(\ref{eq:deltalambda_b}) and (\ref{eq:deltalambda}) from the plane wave model from the preceeding section, which are observed to be in good agreement with the results of the simulations. While the broadband mirror cavity linewidth increases as $1/l$, as the length is reduced, the Fano cavity linewidth saturates to a value determined solely by the Fano mirror linewidth and the internal losses (Eq.~\ref{eq:deltalambda_f})).

In a second series of simulations the finite size of the grating as well the Gaussian profile of the incident beam are taken into account by assuming a grating with width $W=3w_0$, where $w_0$ is the Gaussian beam waist (radius) focussed on the grating. Perfectly matched layers are used in the FEM simulations. In order to get tractable simulations as well as to emphasize finite-size and collimation effects, a waist of 30 $\mu$m---substantially smaller than that used in the experiments (see Sec.~\ref{sec:experiment})---is assumed. With such a waist the reflectivity of the grating at resonance is $\sim 96$\%, the grating being now assumed nonabsorptive ($n=2$). This ensures that the total cavity losses are still $\simeq 8\%$. The broadband cavity consists of the same Bragg grating mirrors as previously, but with a width $W$, and is illuminated by the same Gaussian beam. The variation of the cavity linewidths with the length for both cavities are shown in Fig.~\ref{fig:comsol} by the blue and red dots and are essentially observed not to significantly differ from the plane wave/infinite structure simulation results, except for the broadband mirror cavity in the long cavity regime, where a slight broadening is observed for the chosen waist of 30 $\mu$m. This broadening can be alternatively estimated using a plane wave decomposition of the incident Gaussian beam and using the planar Fabry-Perot spatiospectral transfer function~\cite{Lee2002} (see Appendix). The results of such an estimation for a Gaussian beam waist of 30 $\mu$m (empty blue circles)  are observed to be in very good agreement with the FEM simulation results. For a larger waist of 90 $\mu$m (empty blue squares), comparable to that used in the experiments with grating $A$ discussed in Sec.~\ref{sec:experiment}, collimation effects are predicted to be negligible in the range of cavity lengths considered.


\section{Experimental methods and results}
\label{sec:experiment}

\subsection{Large-area, high-Q suspended subwavelength gratings}

\begin{figure}[h]
\includegraphics[width=\columnwidth]{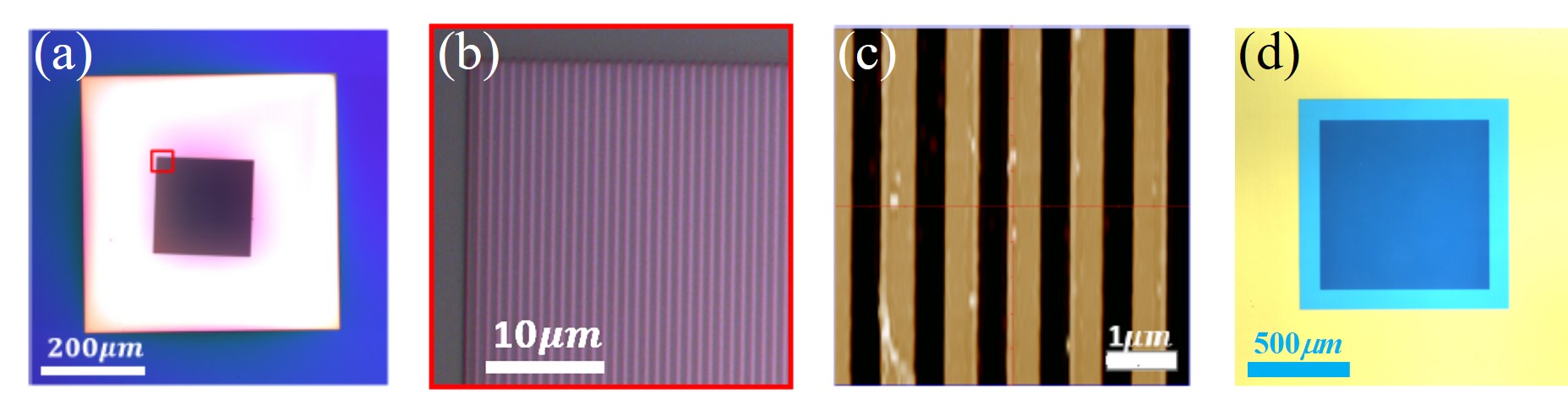}\\
\includegraphics[width=\columnwidth]{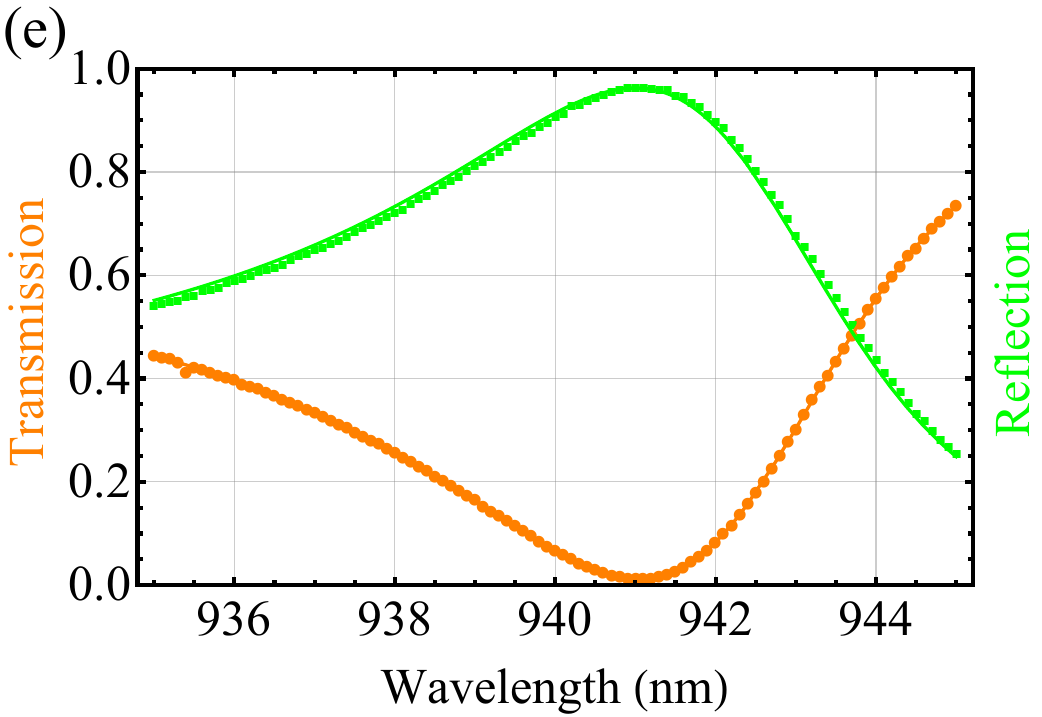}\\
\includegraphics[width=\columnwidth]{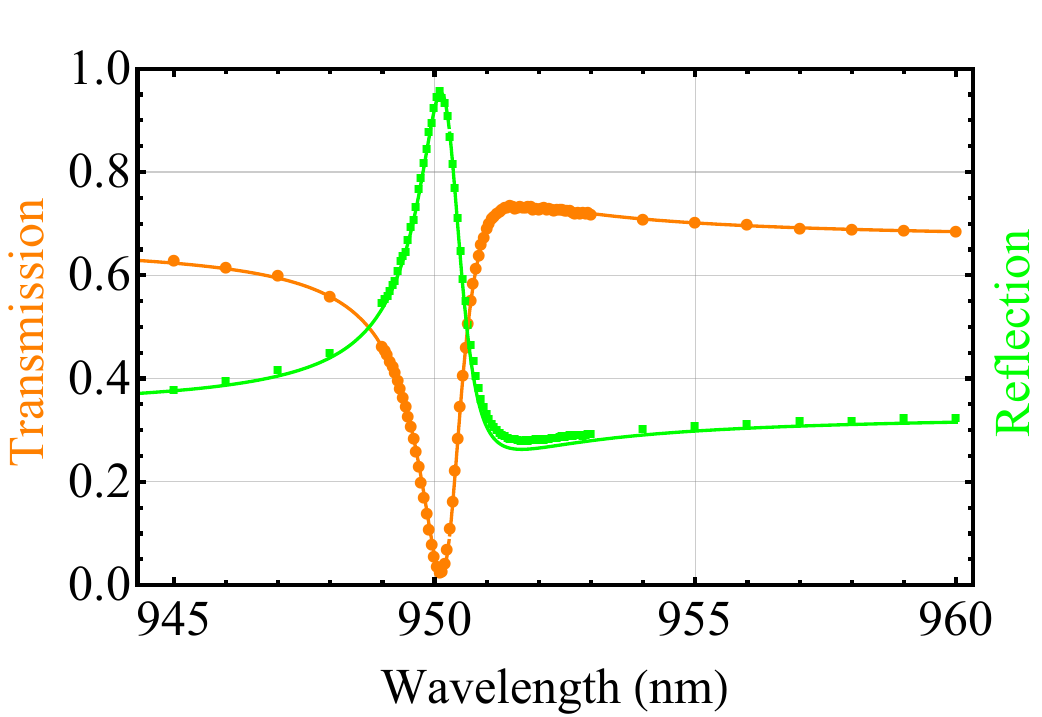}
\caption{(a) Photograph of $A$ grating membrane. (b) Zoom image of the red square in (a). (c) AFM scan image of grating $A$'s fingers. (d) Photograph of $B$ grating membrane. (e) Normalized transmission and reflectivity spectra of grating $A$. The dashed and dot-dashed lines show the results of fits with the lossy grating model. (f) Same for grating $B$.}
\label{fig:gratings}
\end{figure}

In this section and the next we report on experiments carried out with two similar gratings, differing essentially by the optical quality factor of the Fano resonance. Both gratings were realized on highly-pretensioned, suspended silicon nitride films made by Norcada~\cite{Norcada}. The first grating ($A$) was patterned on a high tensile stress ($\sim$ GPa), 203 nm-thick, 500 $\mu$m-square stoichiometric SiN film suspended on a 500 $\mu$m-thick, 5 mm-square silicon frame using electron beam lithography and chemical etching following the recipe of Ref.~\cite{Nair2017}. The grating is 200 $\mu$m$\times$200 $\mu$m and has trapezoidal fingers with period 848 nm, mean finger width 349 nm and depth 114 nm, as determined by AFM profilometry~\cite{Darki2021}. The refractive index at the resonance wavelegnth is 2.0. The normalized transmission and reflectivity spectra of grating $A$ under normal incidence illumination with a TM-polarized Gaussian beam with a waist of 90 $\mu$m are shown in Fig.~\ref{fig:gratings}(a). The grating exhibits a high-reflectivity resonance at 941 nm, with a minimum transmission of 1.3 \% and a maximum reflectivity of 96.0 \%. The resonance HWHM is 3.1 nm, corresponding to a $Q$ of 153. The beam waist was chosen so as to maximize the resonant reflectivity level by minimizing the collimation and finite-size effects~\cite{Magnusson1993,Jacob2000,Bendickson2001} discussed in detail in Ref.~\cite{ToftVandborg2021}.

The second grating ($B$) was made by Norcada on a low stress, 156 nm-thick, 1 mm-square SiN film suspended on a 500 $\mu$m-thick, 5 mm-square silicon frame. The grating is 800 $\mu$m$\times$800 $\mu$m and has fingers with period 850 nm, mean finger width 650 nm and depth 46 nm. The refractive index is 2.14 in the relevant wavelength range. The lower thickness of the $B$ grating film allowed for decreasing the coupling out of the guided mode, and thus increasing the $Q$ factor of the resonance. As the coupling with the guided mode is decreased, though, the beam size needs to be concomittantly increased, as discussed in~\cite{ToftVandborg2021}. Given its much larger size grating $B$ allows for operating with a beam waist of 300 $\mu$m and thereby achieve a similar maximum reflectivity of 95\% at a resonant wavelength of 950.5 nm, while reducing the resonance linewidth by a factor $\sim$6 (Fig.~\ref{fig:gratings}(b)) and resulting in a $Q$-factor of 1010.

The best fit parameter results of the spectra of Fig.~\ref{fig:gratings} with the model of Sec.~\ref{sec:theory} are summarized in Table~\ref{tab:gratings}.

\begin{table}
\caption{Best fit parameter results of the spectra of Fig.~\ref{fig:gratings} with the lossy grating model of Sec.~\ref{sec:theory}.}
  \label{tab:gratings}
  \begin{tabular}{|c|cccccc|}
   \hline
   Grating & $\lambda_0$ (nm) & $\lambda_1$ (nm) & $\gamma_\lambda$ (nm) & $|t_d|^2$ & $\beta$ ($\mu$m$^{-1}$) & $L$\\\hline
   $A$ & 941.0 & 942.4 & 3.1 & 76.6\% & 3.1 & 2.7\% \\
   $B$ & 950.13 & 950.28 & 0.47 & 66.5\% & 0.63 & 2.5\% \\\hline
  \end{tabular}
\end{table}

\subsection{Experimental cavity setup}

\begin{figure}[h]
\includegraphics[width=\columnwidth]{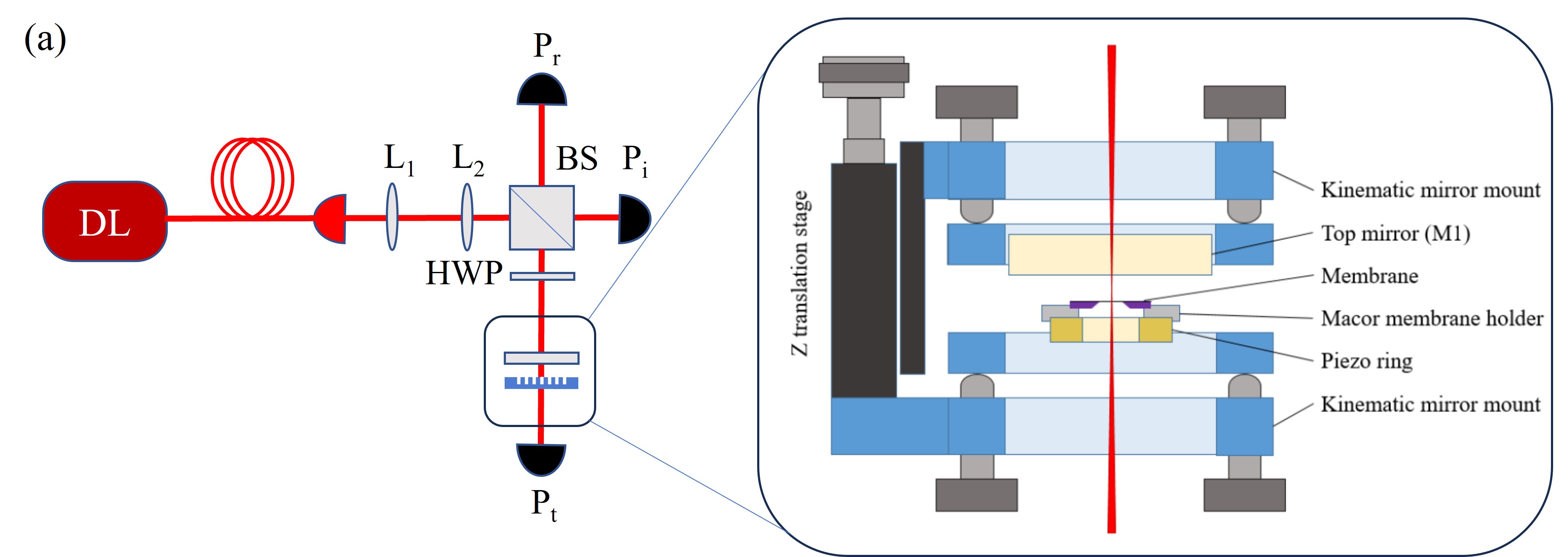}\\
\begin{subfigure}{0.49\columnwidth}
\includegraphics[width=\columnwidth]{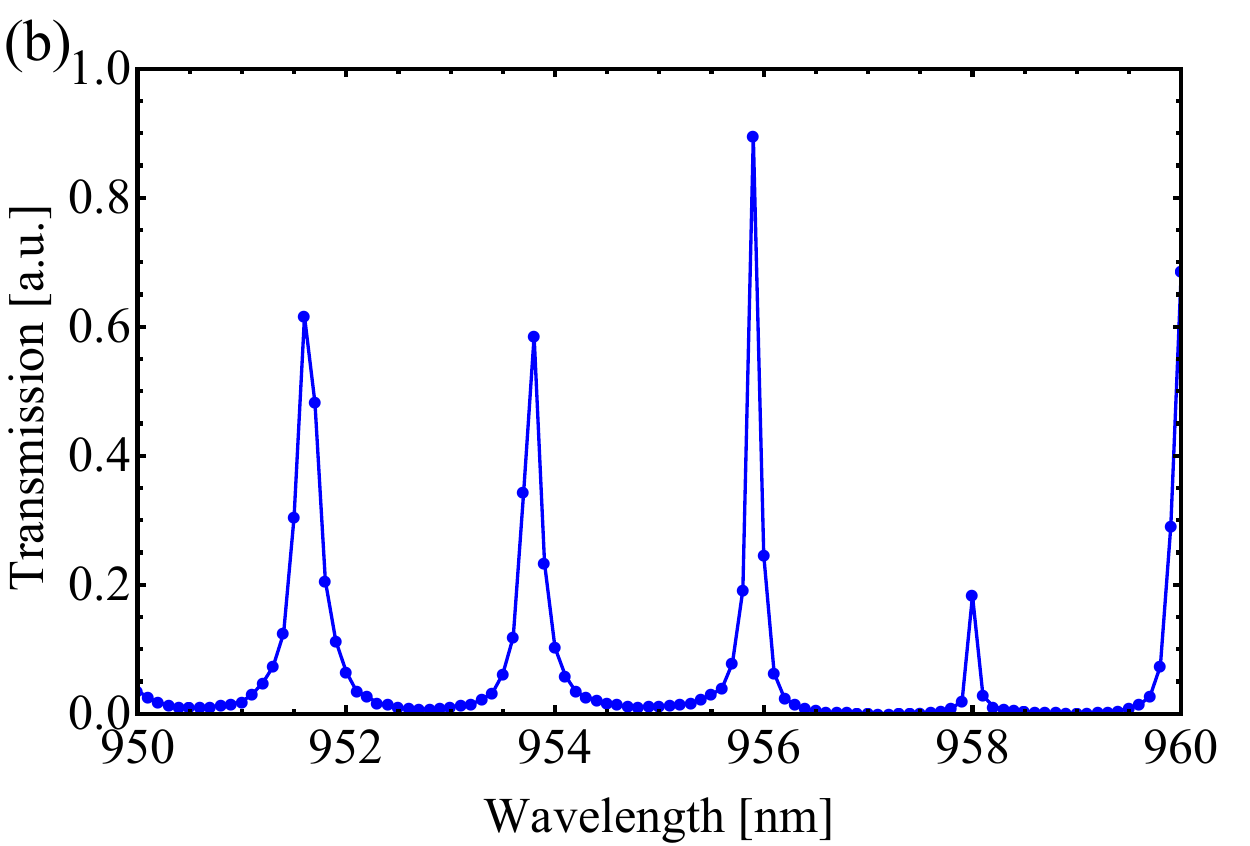}
\end{subfigure}
\begin{subfigure}{0.49\columnwidth}
\includegraphics[width=\columnwidth]{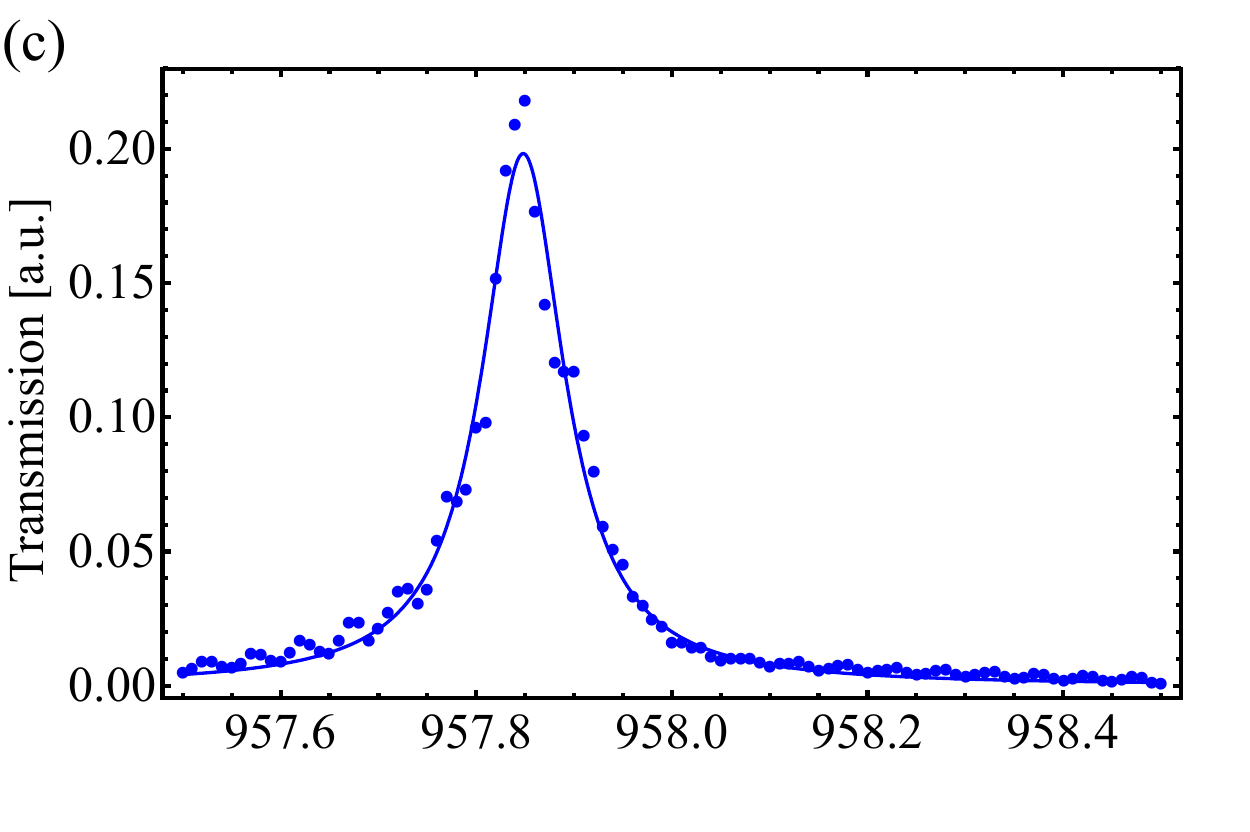}
\end{subfigure}
\caption{(a) Schematic of the experimental setup. (b),(c) Examples of measured broad (b) and narrow (c) range cavity transmission spectra for a broadband cavity with losses equivalent to that of the Fano cavity with grating $A$, yielding an FSR of ($2.1\pm0.1$) nm and a cavity HWHM of ($51.1\pm 1.3$) pm.}
\label{fig:setup}
\end{figure}

The grating/cavity transmission measurements were performed using the setup shown in Fig.~\ref{fig:setup}. After spatial filtering in a single mode fiber, linearly polarized monochromatic light from a tunable diode laser (Toptica DLC Pro) is focused on the grating/cavity. The light transmitted and reflected by the grating/cavity is detected by photodiodes $P_t$ and $P_r$, respectively, referenced to the input power $P_r$ measured in transmission of the incoupling beamsplitter (BS).

For the grating transmission/reflectivity measurements of Fig.~\ref{fig:gratings} the top mirror is removed. For the cavity measurements two configurations are investigated: \textit{(i)} the Fano cavity configuration---consisting of one of the gratings and the partially reflecting top mirror $M_1$, whose reflectivity smoothly varies in the wavelength range considered, thus allowing to adjust the total loss of each cavity---and \textit{(ii)} the broadband cavity configuration---consisting of mirror $M_1$ and a broadband highly reflective (99.8\%) mirror.

The cavity setup consists of three adjustment mounts; two kinematic mirror mounts (Thorlabs Polaris-K13SP) are used to align the two reflective elements (grating/top mirror or HR/top mirrors), and a linear translation stage (Thorlabs NFL5DP20) is used to vary the distance between them. The linear stage is moved by a differential adjuster with a resolution of 50 $\mu$m per revolution that allows for varying the cavity length smoothly so as to be able to count the individual interference fringes. The mounts are also equipped by piezoactuators on top of their adjusters allowing for controlling their position with sub-micron resolution. The whole setup is installed on XYZ translation platform that is used to center the grating with respect to the incidence beam. A piezo ring actuator (Noliac NAC2124) below the grating membrane is used in the alignment phase to scan the cavity length and determine the length at which the cavity resonant wavelength matches the grating resonant wavelength $\lambda_0$.

Once the distance between the cavity mirrors is fixed, broad and narrow wavelength scans are performed and the cavity transmission recorded. The broad range scans allow for measuring the cavity Free Spectral Range (FSR) and calculating the cavity length, taking into account the wavelength dependence of the mirrors. The narrow range scans allow for determining the cavity linewidth at the said cavity length. An example of such scans is given in Figs.~\ref{fig:setup}(b) and (c). For short cavity lengths for which only one cavity resonance can be captured in the available laser wavelength tuning range, the cavity length is inferred by counting the number of cavity resonance fringes observed when reducing the cavity length.

\subsection{Experimental results}

\begin{figure*}[h]
\begin{subfigure}{0.49\textwidth}
\includegraphics[width=\columnwidth]{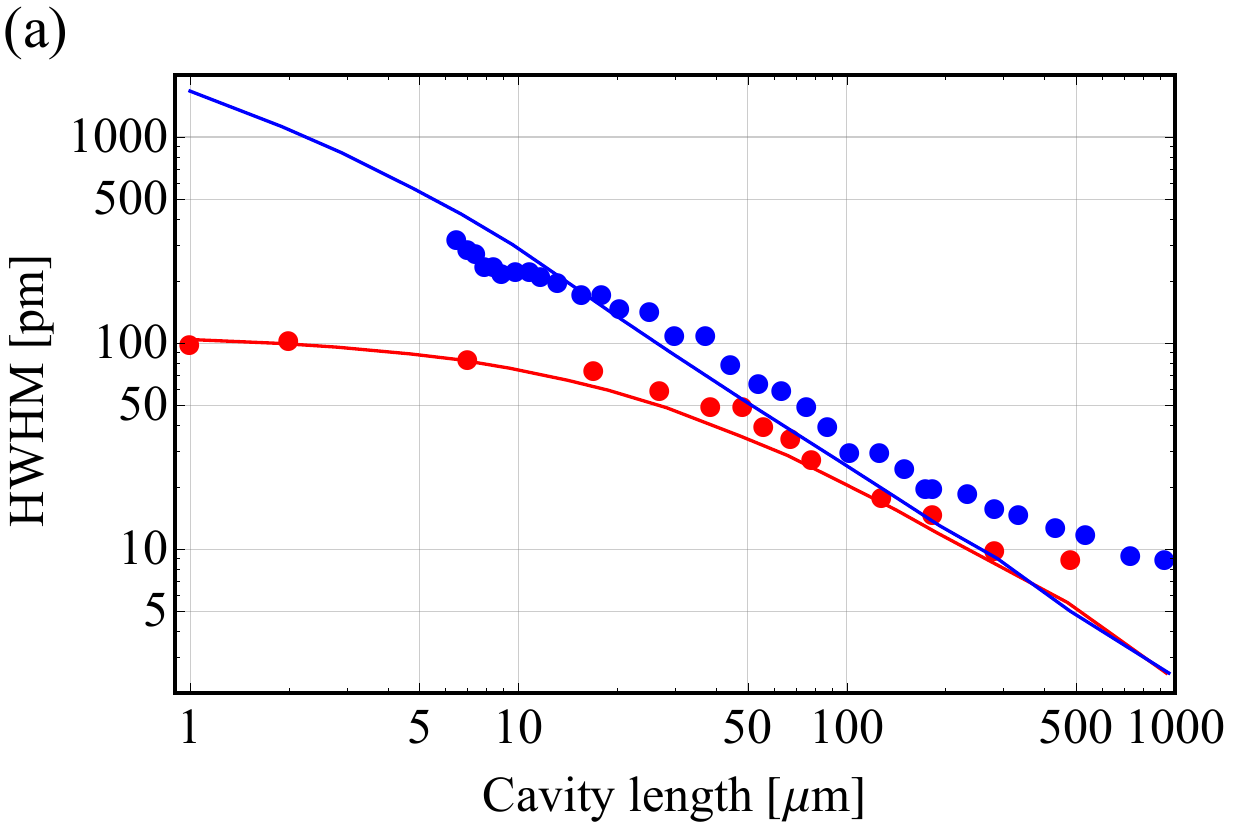}
\end{subfigure}
\begin{subfigure}{0.49\textwidth}
\includegraphics[width=\columnwidth]{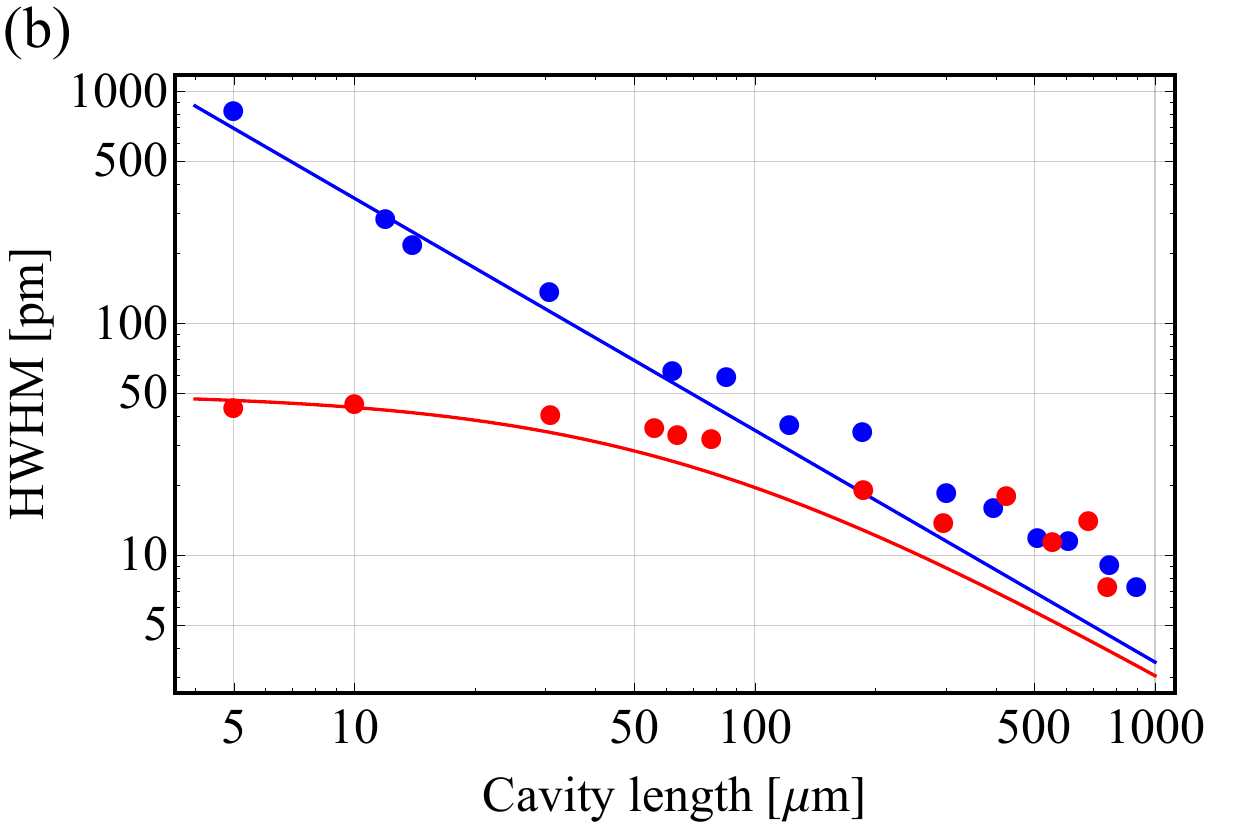}
\end{subfigure}
\begin{subfigure}{0.49\textwidth}
\includegraphics[width=\columnwidth]{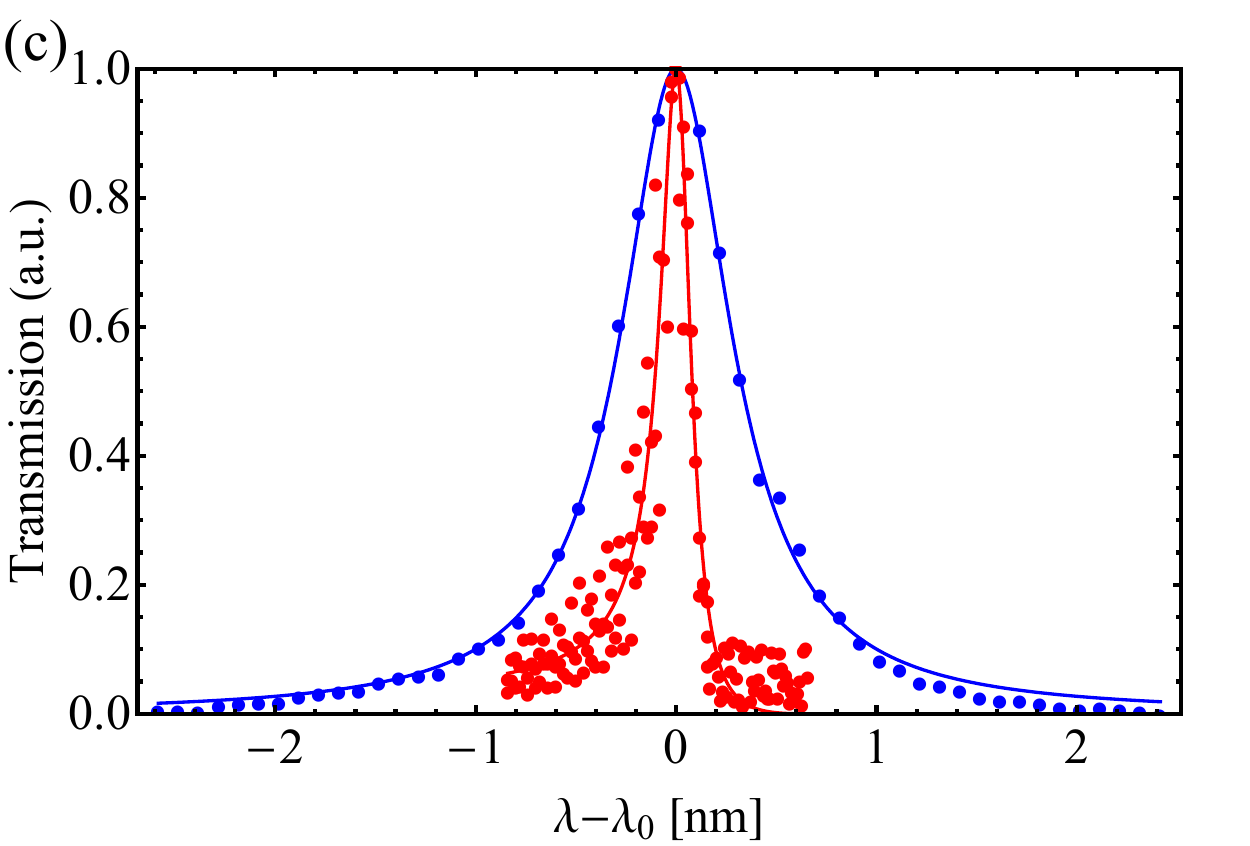}
\end{subfigure}
\begin{subfigure}{0.49\textwidth}
\includegraphics[width=\columnwidth]{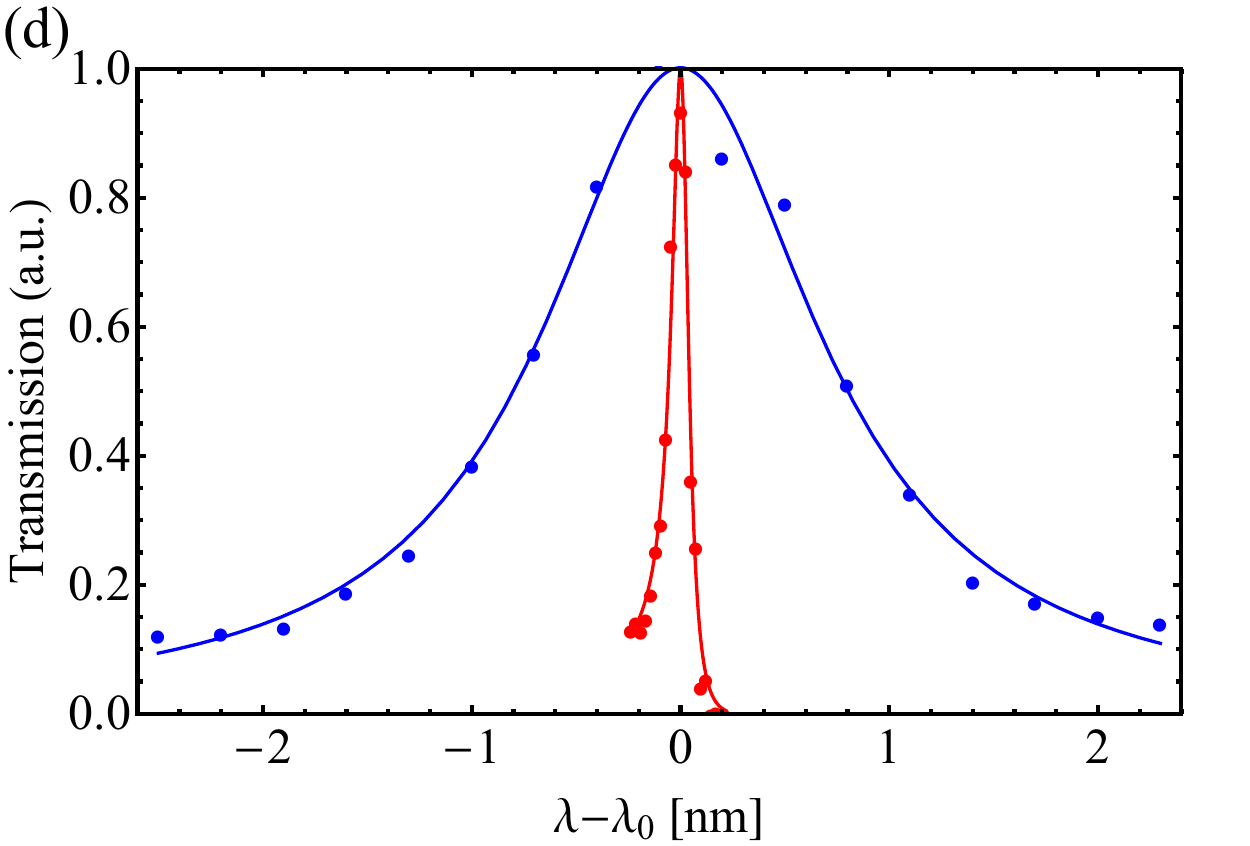}
\end{subfigure}
\caption{(a) Variation of the cavity linewidth with cavity length for the Fano cavity realized with grating $A$ (red points) and the corresponding broadband cavity (blue points). The solid lines show the analytical "plane-wave" cavity predictions. (b) Same for the Fano cavity realized with grating $B$. (c) Measured cavity transmission spectrum for Fano cavity $A$ (red points) and corresponding broadband cavity (blue points) at a cavity length of 6 $\mu$m. The solid lines show the results of fits with Eq.~(\ref{eq:cavitylinewidth}), yielding $\delta\lambda=(94.4\pm4.2)$ pm for the Fano cavity and $(333.9\pm4.6)$ pm for the broadband cavity. (d) Same for Fano cavity $B$ and a cavity length of 5 $\mu$m. The linewidths resulting from the fit are $(43.5\pm2.9)$ pm for the Fano cavity and $(799\pm55)$ pm for the broadband cavity.}
\label{fig:linewidthvsL}
\end{figure*}

The variations of the cavity linewidth as a function of cavity length are shown in Fig.~\ref{fig:linewidthvsL} for the Fano cavities realized with gratings $A$ and $B$, respectively. For each Fano cavity, the variations of the corresponding broadband cavity with the same total losses (8\% for Fano cavity $A$ and 10\% for Fano cavity $B$) are also shown. The different scaling of the linewidth with the length of the two types of cavity are clearly observed and the measured linewidths are in overall good agreement with the "plane-wave" cavity analytical predictions of (\ref{eq:deltalambda}) and (\ref{eq:deltalambda_b}) for cavity lengths below $\sim100$ $\mu$m. The observed HWHM of Fano cavity $A$ at a length $l\sim 1$ $\mu$m is 100 pm, corresponding to an optical Q-factor of $Q=\lambda_0/(2\delta \lambda)\sim 4700$, an enhancement of a factor 16.7 with respect to the corresponding broadband cavity. The HWHM of Fano cavity $B$ at the shortest achieved length $l\sim 5$ $\mu$m is 45 pm, corresponding to an optical Q-factor of $10560$, which represents an enhancement of a factor 15.4 with respect to the corresponding broadband cavity with the same length.

At longer cavity lengths the observed linewidths of both types of cavity are somewhat systematically broadened with respect to the theoretical expectations. As linewidth broadening due to collimation effects is expected to be negligible for the large waists used in the experiments, we attribute the observed broadening to imperfect parallelism between the cavity mirrors and/or small drifts in cavity length (which was not actively stabilized in these experiments) during the scans, both effects being expected to be more important for longer cavities~\cite{Naesby2018}.

The good agreement between the observed and predicted linewidths at short lengths supports the robustness of ultrashort Fano cavities with respect to imperfect parallelism. Let us note, however, that in order to achieve very short ($\sim\mu$m) cavity lengths, a very good control of the parallelism is still required, as any contact of the frame surface with the mirror may physically limit the cavity length. Here, the much smaller lateral size of the membrane chip (5 mm) than the standard HR mirror (25 mm) made this more easily achievable in the Fano cavity configuration than in the broadband cavity configuration. We thus expect that using smaller mirrors or a membrane sandwich configuration using spacers~\cite{Salimi2023} should facilitate the realization of few-micron---or even sub-micron as demonstrated in Ref.~\cite{Salimi2023}---Fano cavities. One can also envisage piezoelectric fine tuning of these Fano cavity as demonstrated in~\cite{Nair2019}. This would facilitate for instance the systematic study of the different optomechanical effects at play in ultrashort Fano microcavity~\cite{Manjeshwar2023} and the investigations of bound state in the continuum resonance in coupled Fano mirror systems~\cite{Fitzgerald2021,Peralle2023}. While the latter would represent a way of potentially reducing further the cavity linewidth, an alternative could be achieved by using gratings with higher Q Fano resonances together with increased peak reflectivity~\cite{Singh2023}.


\section{Conclusion}
\label{sec:conclusion}

By using large area, highly reflective and high-Q resonant subwavelength silicon nitride gratings suspended on top of a conventional broadband mirror we realized Fano microcavities with a spectral sensitivity substantially enhanced as compared to that of the grating alone or that of the equivalent broadband mirror cavity. In the few micron-length regime optical Qs in the 5000-10000 range are obtained, larger by more than one order of magnitude than those of the corresponding broadband cavities or of the Fano mirrors. The experimental observations are in good agreement with a "plane-wave" Fano cavity model and finite element simulations taking the finite size of the grating/cavity and the Gaussian nature of the incident beam into account allow for assessing collimation and finite-size linewidth broadening.

These results are promising for a number of optomechanics applications, e.g. the realization of low-light level optical bistable and nonreciprocal optomechanical cavities~\cite{Sang2022,Xu2022,Zhou2023,Enzian2023}, the investigations of Fano cavity optomechanics~\cite{Cernotik2019,Fitzgerald2021,Manjeshwar2023,Peralle2023}  and multiple-membrane cavity optomechanics~\cite{Xuereb2012,Xuereb2014,Nair2017,Gartner2018,Piergentili2018,Wei2019,Manjeshwar2020,Yang2020b} in the strong coupling regime, but also potentially for optical sensing~\cite{Naserbakht2019,AlSumaidae2021,Hornig2022,Salimi2023} and Fano lasers~\cite{Mork2014,Yu2017}.


\section{Appendix: Spatiospectral decomposition}
\label{sec:appendix}

In Sec.~\ref{sec:comparison} we make use of a one-dimensional plane wave decomposition of the incident field~\cite{ToftVandborg2021} and introduce the Fourier transform of the field amplitude in real space $E_\textrm{in}(x)$ as
\begin{equation}
\tilde{E}_\textrm{in}(k_x)=\int_{-\infty}^{\infty} dx\,E_\textrm{in}(x)e^{-ik_xx},
\end{equation}
where $x$ is the spatial coordinate in the direction perpendicular to the grating fingers and $k_x$ is the spatial frequency coordinate. The angular spectrum of the transmitted field by the cavity is obtained by multiplying the angular spectrum of the incident field by the planar Fabry-Perot interferometer transfer function $t_\textrm{FP}(k_x)$
\begin{equation}
E_\textrm{tr}(x)=\frac{1}{2\pi}\int_{-\infty}^{\infty}dk_x\, t_\textrm{FP}(k_x)\tilde{E}_\textrm{in}(k_x)e^{ik_xx},
\end{equation}
where, up to an irrelevant phase factor, $t_\textrm{FP}(k_x)$ is given by~\cite{Lee2002}
\begin{equation}
t_\textrm{FP}(k_x)=\frac{t_m^2}{1-r_m^2e^{2iL\sqrt{k^2-k_x^2}}}.
\end{equation}
For an incident Gaussian beam amplitude $E_\textrm{in}(x)=E_0e^{-x^2/w_0^2}$, one has
\begin{equation}
\tilde{E}_\textrm{in}=E_0\sqrt{\pi}w_0 e^{-(k_xw_0/2)^2}.
\end{equation}
The normalized transmission of the broadband cavity under Gaussian beam illumination is then given by
\begin{equation}
\mathcal{T}=\frac{\int_{-\infty}^{\infty}dk_x\,|t_\textrm{FP}(k_x)\tilde{E}_\textrm{in}(k_x)|^2}{\int_{-\infty}^{\infty}dk_x\,|\tilde{E}_\textrm{in}(k_x)|^2}.
\end{equation}

\section*{Funding}
Independent Research Fund Denmark, Novo Nordisk Fonden.

\section*{Acknowledgments}
We are grateful to Norcada for their assistance with the design and fabrication of the samples used in this study.

\section*{Disclosures}
The authors declare no conflicts of interest.

\section*{Data Availability Statement}
Data underlying the results presented in this paper are not publicly available at this time but may be obtained from the authors upon reasonable request.




\begin{thebibliography}{99}

\bibitem{Vahala2003} K. J. Vahala, "Optical microcavities," Nature {\bf 424}, 839 (2003).

\bibitem{Aspelmeyer2014} Markus Aspelmeyer, Tobias J. Kippenberg, and Florian Marquardt, "Cavity optomechanics," Rev. Mod. Phys. {\bf 86}, 1391 (2014).

\bibitem{Naserbakht2019} S. Naserbakht and A. Dantan, "Squeeze film pressure sensors based on SiN membrane sandwiches," Sens. Actuators A: Phys. {\bf 298}, 111588 (2019).
\bibitem{AlSumaidae2021} S. Al-Sumaidae, L. Bu, G. J. Hornig, M. H. Bitarafan, R. G. DeCorby, "Pressure sensing with high-finesse monolithic buckled-dome microcavities," Appl. Opt. {\bf 60}, 9219-9224 (2021).
\bibitem{Hornig2022} G. J. Hornig, K. G. Scheuer, E. B. Dew, R. Zemp, and R. G. DeCorby, "Ultrasound sensing at thermomechanical limits with optomechanical buckled-dome microcavities," Opt. Express {\bf 30}, 33083-33096 (2022).
\bibitem{Salimi2023} M. Salimi, R. V. Nielsen, H. B. Pedersen, and A. Dantan, "Squeeze film absolute pressure sensors with sub-millipascal sensitivity," arxiv:2312.11915 (2023).


\bibitem{Kemiktarak2012} U. Kemiktarak, M. Metcalfe, M. Durand, and J. Lawall, "Mechanically compliant grating reflectors for optomechanics," Appl. Phys. Lett. {\bf 100}, 061124 (2012).
\bibitem{Bui2012} C. H. Bui, J. Zheng, S. W. Hoch, L. Y. T. Lee, J. G. E. Harris, and C. W. Wong, "High-reflectivity, high-Q micromechanical membranes via guided resonances for enhanced optomechanical coupling," Appl. Phys. Lett. {\bf 100}, 021110 (2012).
\bibitem{Kemiktarak2012a} U. Kemiktarak,  M. Durand, M. Metcalfe, and J. Lawall, "Cavity optomechanics with sub-wavelength grating mirrors," New J. Phys. {\bf 14}, 125010 (2012).
\bibitem{Norte2016} R. A. Norte, J. P. Moura, and S. Gr\"{o}blacher, "Mechanical Resonators for Quantum Optomechanics Experiments at Room Temperature," Phys. Rev. Lett. {\bf 116}, 147202 (2016).
\bibitem{Reinhardt2016} C. Reinhardt, T. M\"{u}ller, A. Bourassa, and J. C. Sankey, "Ultralow-Noise SiN Trampoline Resonators for Sensing and Optomechanics," Phys. Rev. X {\bf 6}, 021001 (2016).
\bibitem{Chen2017}  X. Chen, C. Chardin, K. Makles, C. Ca\"{e}r, S. Chua, R. Braive, I. Robert-Philip, T. Briant, P.-F. Cohadon, A. Heidmann, T. Jacqmin, and S. Deleglise, "High-finesse Fabry-Perot cavities with bidimensional Si3N4 photonic-crystal slabs," Light Sci. Appl. {\bf 6}, e16190 (2017).
\bibitem{Zhou2023} F. Zhou, Y. Bao, J. J. Gorman, and J. R. Lawall, "Cavity Optomechanical Bistability with an Ultrahigh Reflectivity Photonic Crystal Membrane," Laser Photonics Rev. 2300008 (2023).
\bibitem{Enzian2023} G. Enzian, Z. Wang, A. Simonsen, J. Mathiassen, T. Vibel, Y. Tsaturyan, A. Tagantsev, A. Schliesser, and E. S. Polzik, "Phononically shielded photonic-crystal mirror membranes for cavity quantum optomechanics," Optics Express {\bf 31}, 13040-13052 (2023).

\bibitem{Naesby2018} A. Naesby and A. Dantan, "Microcavities with suspended subwavelength structured mirrors," Opt. Express {\bf 26}, 29886-29894 (2018).

\bibitem{Wang1993} S. Wang and R. Magnusson, "Theory and applications of guided-mode resonance filters," Appl. Opt. {\bf 32}, 2606-2613 (1993).
\bibitem{Fan2003} S. Fan, W. Suh, and J. D. Joannopoulos, "Temporal coupled-mode theory for the Fano resonance in optical resonators," J. Opt. Soc. Am. A {\bf 20}, 569-572 (2003).
\bibitem{Limonov2017} M. F. Limonov, M. V. Rybin, A. N. Poddubny, and Y. S. Kivshar, "Fano resonances in photonics," Nat. Photo. {\bf 11}, 543-554 (2017).
\bibitem{Quaranta2018} G. Quaranta, G. Basset, O. J. F. Martin, and B. Gallinet, "Recent advances in resonant waveguide gratings," Laser Photonics Rev. {\bf 12}, 1800017 (2018).


\bibitem{Cernotik2019} A. Cernotik, A. Dantan, and C. Genes, "Cavity quantum electrodynamics with frequency-dependent reflectors," Phys. Rev. Lett. {\bf 122}, 243601 (2019).
\bibitem{Fitzgerald2021} J. M. Fitzgerald, S. K. Manjeshwar, W. Wieczorek, and P. Tassin, "Cavity optomechanics with photonic bound states in the continuum," Phys. Rev. Res. {\bf 3}, 013131 (2021).
\bibitem{Peralle2023} C. P{\'e}ralle, S. K. Manjeshwar, A. Ciers, W. Wieczorek, and P. Tassin, "Quasi bound states in the continuum in photonic-crystal-based optomechanical microcavities," arxiv:2306.17831 (2023).
\bibitem{Manjeshwar2023} S. K. Manjeshwar, A. Ciers, J. Monsel, H. Pfeifer, C. Peralle, S. M. Wang, P. Tassin, and W. Wieczorek, "Integrated microcavity optomechanics with a suspended photonic crystal mirror above a distributed Bragg reflector," Opt. Express {\bf 31}, 30212-30226 (2023).

\bibitem{Ossiander2023} M. Ossiander, M. L. Meretska, S. Rourke, C. M. Spaegele, X. Yin, I. C. Benea-Chelmus, and F. Capasso, "Metasurface-stabilized optical microcavities," Nat. Commun. {\bf 14}, 114 (2023).


\bibitem{Sang2022} Y. Sang, J. Xu, K. Liu, W. Chen, Y. Xiao, Z. Zhu, N. Liu, and J. Zhang, "Spatial Nonreciprocal Transmission and Optical Bistabilty Based on Millimiter-scale Suspended Metasurface," Adv. Opt. Mat. 2201523 (2022).
\bibitem{Xu2022} J. Xu, K. Liu, Y. Sang, Z. Tan, C. Guo, and Z. Zhu, "Millimiter-scale ultrathin suspended metasurface integrated high-finesse optomechanical cavity," Opt. Lett. {\bf 47}/, 5481-5484 (2022).



\bibitem{Xuereb2012} A. Xuereb, C. Genes, and A. Dantan, "Strong Coupling and Long-Range Collective Interactions in Optomechanical Arrays," Phys. Rev. Lett. {\bf 109}, 223601 (2012).
\bibitem{Xuereb2014} A. Xuereb, C. Genes, A. Pupillo, M. Paternostro, and A. Dantan, "Reconfigurable Long-Range Phonon Dynamics in Optomechanical Arrays," Phys. Rev. Lett. {\bf 112}, 133604 (2014).
\bibitem{Nair2017} B. Nair, A. Naesby, and A. Dantan, "Optomechanical characterization of silicon nitride membrane arrays," Opt. Lett. {\bf 42}, 1341-1344 (2017).
\bibitem{Gartner2018} C. G\"{a}rtner, J. P. Moura, W. Haaxman, R. A. Norte, and S. Gr\"{o}blacher, "Integrated optomechanical arrays of two high reflectivity SiN membranes," Nano Lett. {\bf 18}, 7171-7175 (2018).
\bibitem{Piergentili2018} P. Piergentili, L. Catalini, M. Bawaj, S. Zippilli, N. Malossi, R. Natali, D. Vitali, and G. D. Giuseppe, "Two-membrane cavity optomechanics," New J. Phys. {\bf 20}, 083024 (2018).
\bibitem{Wei2019} X. Wei, J. Sheng, C. Yang, Y. Wu, and H. Wu, "Controllable two-membrane-in-the-middle cavity optomechanical system," Phys. Rev. A {\bf 99}, 023851 (2019).
\bibitem{Manjeshwar2020} S. K. Manjeshwar, K. Elkhouly, J. M. Fitzgerald, M. Ekman, Y. Zhang, F. Zhang, S. M. Wang, P. Tassin, and W. Wieczorek, "Suspended photonic crystal membranes in AlGaAs heterostructures for integrated multi-element optomechanics," Appl. Phys. Lett. {\bf 116}, 264001 (2020).
\bibitem{Yang2020b} C. Yang, X. Wei, J. Sheng, and H. Wu, "Phonon heat transport in cavity-mediated nanomechanical resonators," Nat. Comm. {\bf 11}, 4626 (2020).

\bibitem{Mork2014} J. M{\o}rk, Y. Chen, and M. Heuck, “Photonic crystal Fano laser: terahertz modulation and ultrashort pulse generation,” Phys. Rev. Lett. {\bf 113},
163901 (2014).
\bibitem{Yu2017} Y. Yu, W. Xue, E. Semenova, K. Yvind, and J. M\o rk, "Demonstration of a self-pulsing photonic crystal Fano laser," Nat. Photon. {\bf 11}, 81–84 (2017).


\bibitem{Darki2022} A. A. Darki, R. V. Nielsen, J. V. Nygaard, and A. Dantan, "Mechanical investigations of free-standing SiN membranes patterned with one-dimensional photonic crystal structures", J. Appl. Phys. {\bf 131}, 195901 (2022).


\bibitem{Popov1986} E. Popov, L. Mashev, and D. Maystre, "Theoretical Study of the Anomalies of Coated Dielectric Gratings," Optica Acta: Int. J. Opt. {\bf 33}, 607-619 (1986).
\bibitem{Bykov2015} D. A. Bykov and L. L. Doskolovich, "Spatiotemporal coupled-mode theory of guided-mode resonant gratings," Opt. Express {\bf 23}, 19234-19241 (2015).
\bibitem{Darki2021} A. A. Darki, A. Parthenopoulos, J. V. Nygaard, and A. Dantan, "Profilometry and stress analysis of suspended nanostructured thin films," J. Appl. Phys. {\bf 129}, 065302 (2021).
\bibitem{Parthenopoulos2021} A. Parthenopoulos, A. A. Darki, B. R. Jeppesen, and A. Dantan, "Optical spatial differentiation using suspended subwavelength gratings", Opt. Express {\bf 29}, 6481-6494 (2021).


\bibitem{Magnusson1993} R. Magnusson and S. S. Wang, ‘‘Optical waveguide-grating filters,’’ in International Conference on Holography, Correlation Optics, and Recording Materials, O. V. Angelsky, ed., Proc. SPIE {\bf 2108}, 380–390 (1993).
\bibitem{Jacob2000} D. K. Jacob, S. C. Dunn, and M. G. Moharam, "Design considerations for narrow-band dielectric resonant grating reflection filters of finite length", J. Opt. Soc. Am. A {\bf 17}, 1241-1249 (2000).
\bibitem{Bendickson2001} J. M. Bendickson, E. N. Glytsis, T. K. Gaylord, and D. L. Brundrett, "Guided-mode resonant subwavelength gratings: effects of finite beams and finite gratings,"  J. Opt. Soc. Am. A {\bf 18}, 1912-1928 (2001).
\bibitem{ToftVandborg2021} C. Toft-Vandborg, A. Parthenopoulos, A. A. Darki, and A. Dantan, "Collimation and finite-size effects in suspended resonant guided-mode gratings", J. Opt. Soc. Am. A {\bf 38}, 1714-1725 (2021).

\bibitem{Lee2002} J. Y. Lee, J. W. Hahn, and H.-W. Lee, "Spatiospectral transmission of a plane-mirror Fabry-Perot interferometer with nonuniform finite-size diffraction beam illuminations," J. Opt. Soc. Am. A {\bf 19}, 973-984 (2002).


\bibitem{Norcada} Norcada, Inc. (www.norcada.com).
\bibitem{MIST} T. Germer, {\it Modeled Integrated Scatter Tool}, available at http://physics.nist.gov/scatmech.

\bibitem{Nair2019} B. Nair, A. Naesby, B. R. Jeppesen, and A. Dantan, "Suspended silicon nitride thin films with enhanced and electrically tunable reflectivity," Phys. Scr. {\bf 14}, 125013 (2019).

\bibitem{Singh2023} G. Singh, T. Mitra, S. P. Madsen, and A. Dantan, "Highly reflective, high-Q resonant dual-period subwavelength gratings," in preparation (2023).







\end{thebibliography}
\end{document}